\documentclass[twocolumn]{aastex7}
\usepackage{graphicx}
\usepackage{amsmath}
\usepackage{scalerel}
\usepackage{apjfonts}
\usepackage{tikz}
\usetikzlibrary{svg.path}
\usepackage{comment}
\usepackage{array,multirow,color,tablefootnote}
\usepackage{xcolor}

\definecolor{orcidlogocol}{HTML}{A6CE39}
\tikzset{
  orcidlogo/.pic={
    \fill[orcidlogocol] svg{M256,128c0,70.7-57.3,128-128,128C57.3,256,0,198.7,0,128C0,57.3,57.3,0,128,0C198.7,0,256,57.3,256,128z};
    \fill[white] svg{M86.3,186.2H70.9V79.1h15.4v48.4V186.2z}
                 svg{M108.9,79.1h41.6c39.6,0,57,28.3,57,53.6c0,27.5-21.5,53.6-56.8,53.6h-41.8V79.1z M124.3,172.4h24.5c34.9,0,42.9-26.5,42.9-39.7c0-21.5-13.7-39.7-43.7-39.7h-23.7V172.4z}
                 svg{M88.7,56.8c0,5.5-4.5,10.1-10.1,10.1c-5.6,0-10.1-4.6-10.1-10.1c0-5.6,4.5-10.1,10.1-10.1C84.2,46.7,88.7,51.3,88.7,56.8z};
  }
}

\newcommand\orcidicon[1]{\href{https://orcid.org/#1}{\mbox{\scalerel*{
\begin{tikzpicture}[yscale=-1,transform shape]
\pic{orcidlogo};
\end{tikzpicture}
}{|}}}}

\newcommand{\mss}{m\,s$^{-2}$}
\newcommand{\um}{$\mu$m}
\newcommand{\ubar}{$\mu$bar}
\newcommand{\uJy}{$\mu$Jy}

\newcommand{\water}{H$_{2}$O}
\newcommand{\cotwo}{CO$_{2}$}
\newcommand{\methane}{CH$_{4}$}

\newcommand{\nitrogen}{N$_{2}$}

\newcommand{\tkin}{T_{\rm kin}}
\newcommand{\tvib}{T_{\rm vib}}

\begin{document}

\title{Constraints on the Atmospheric Composition of 2002\,XV$_{93}$ from JWST Spectroscopy}

\newcommand{\target}{2002\,XV$_{93}$}

\author[orcid=0000-0001-9665-8429]{Ian~Wong}
\affiliation{Space Telescope Science Institute, 3700 San Martin Drive, Baltimore, MD 21218, USA}
\email{iwong@stsci.edu}

\author[orcid=0000-0001-8541-8550]{Silvia~Protopapa}
\affiliation{Southwest Research Institute, Solar System Science and Exploration Division, Boulder, CO, USA}
\email{silvia.protopapa@swri.org}

\author[orcid=0000-0001-7168-1577]{Emmanuel~Lellouch}
\affiliation{LIRA, Observatoire de Paris, Université PSL, CNRS, Sorbonne Université, Université Paris Cité, 5 place Jules Janssen, 92195 Meudon, France}
\email{Emmanuel.Lellouch@obspm.fr}

\begin{abstract}
The recent detection of an atmosphere surrounding the trans-Neptunian object (TNO) \target\ from stellar occultation measurements has challenged the longstanding view that only the largest TNOs can sustain an atmosphere. Atmospheric refraction modeling of the occultation light curves indicated a surface pressure of 100--200\,nbar, despite \target's relatively small size ($\sim$510\,km in diameter) and weak surface gravity. Together with the detection of methane fluorescence on Makemake, this result suggests that tenuous atmospheres may be more common among TNOs than previously thought. We report JWST/NIRSpec observations acquired before and after the 2024 stellar occultation measurements, obtained with the PRISM and medium-resolution gratings at resolving powers of $\sim$100 and $\sim$1000, respectively. We detect no statistically significant emission features attributed to methane or carbon monoxide gas. By comparing the higher spectral resolution data with synthetic fluorescence models, we report upper limits for the methane and carbon monoxide surface partial pressures of $(3-10)\,\times\,10^{-6}$ and $(5-30)\,\times\,10^{-6}$\,nbar, respectively, substantially below the atmospheric pressure inferred from the occultation measurements. Additionally, we report no evidence of an extended source of either methane gas or refractory material. Our results indicate that the atmospheric interpretation of the occultation measurements may require either a composition dominated by volatile species other than methane and carbon monoxide, with nitrogen and argon as possible candidates, or a methane-dominated atmosphere confined near the surface with a steeply decreasing vertical density profile.
\end{abstract}



\section{Introduction}\label{sec:intro}
\setcounter{footnote}{0}

The trans-Neptunian object (TNO) (612533)\,\target\ is a mid-sized plutino that lies in the 3:2 mean motion resonance with Neptune. A radiometric analysis of thermal infrared measurements from the Spitzer Space Telescope and the Herschel Space Observatory indicated a mean diameter of $\sim$550\,km and a dark surface with an optical geometric albedo of 4\% \citep{Mommert2012}. Infrared spectroscopy of \target\ from the James Webb Space Telescope (JWST) uncovered the presence of water (\water) and carbon dioxide (\cotwo) ices on the surface \citep{PinillaAlonso2025}.  Remarkably, time-series measurements of a stellar occultation event on UT 2024 January 10 displayed a distinctive photometric signature at ingress and egress that is consistent with a thin atmosphere surrounding \target. Analysis of the occultation chords indicated a sky-projected disk diameter of $\sim$510\,km, while atmospheric refraction modeling of the ingress and egress light curves yielded a surface pressure of 100--200\,nbar \citep{Arimatsu2026}.

Until recently, only the largest TNOs were known to host atmospheres. The atmospheres on Pluto and Triton have surface pressures on the order of 10\,\ubar\ and molecular compositions dominated by nitrogen (\nitrogen), with minor contributions from methane (\methane) and carbon monoxide (CO). For detailed reviews of TNO atmospheres, see, for example, \citet{Stern2008}, \citet{Young2020}, and \citet{Young2021}; see also the recent JWST atmospheric study of Pluto by \citet{Lellouch2025}. Spectroscopy of the dwarf planet Makemake obtained by JWST in 2023 revealed characteristic emission features of gaseous \methane\ \citep{Protopapa2025}. Comparisons with fluorescence models suggested that the observed spectral signature may arise from either localized outgassing or a tenuous bound atmosphere; in the atmospheric scenario, the inferred \methane\ surface partial pressure is of order $\sim$10,pbar, 6 orders of magnitude below the corresponding levels on Pluto and Triton \citep{Protopapa2025}. Meanwhile, stellar occultation measurements of other large TNOs---including Eris, Haumea, and Quaoar---have so far not yielded definitive detections of global atmospheres, with surface pressure upper limits spanning 1--100\,nbar \citep{Sicardy2011,Ortiz2017,Proudfoot2025}.


\begin{deluxetable*}{lccclcccccccc}[t]
\tablewidth{0pc}
\setlength{\tabcolsep}{1.2pt}
\tabletypesize{\small}
\renewcommand{\arraystretch}{0.9}
\tablecaption{
    \target\ Observations
    \label{tab:obs}
}
\tablehead{
    Observation &  $\,\quad\quad\quad$ & 
    Disperser/Filter &  $\,\quad\quad\quad$ & 
    Start Time (UT) &  $\,\quad\quad\quad$ & 
    $r_h$ (au)\tablenotemark{\scriptsize \textnormal{a}} &  $\,\quad\quad$ & 
    $\Delta$ (au)\tablenotemark{\scriptsize \textnormal{a}} & $\,\quad\quad$ & 
    $\alpha$ (deg)\tablenotemark{\scriptsize \textnormal{a}} & $\,\quad\quad$ & 
    Exp. Time (s)
}
\startdata
JWST/NIRSpec & & PRISM/CLEAR & & 2022 November 3 13:26 & & 38.13 & & 37.64 & & 1.32 & & 1225 \\ 
Occultation & & --- & & 2024 January 10 13:12 & & 37.99 & & 37.04 & & 0.39 & & --- \\
JWST/NIRSpec & & G395M/F290LP & & 2024 October 8 19:45 & & 37.90 & & 37.86 & & 1.52 & & 23576 \\
\enddata
\textbf{Note.}
\vspace{-0.15cm}\tablenotetext{\textrm{a}}{The variables $r_h$, $\Delta$, and $\alpha$ refer to the heliocentric distance, observer distance, and phase angle of \target\ at the midpoint of the observations, respectively.}
\vspace{-0.6cm}
\end{deluxetable*}

With an estimated surface pressure of 100--200\,nbar, the atmosphere of \target\ is $\sim$100 times thinner than Pluto and Triton's atmospheres, but also $\sim$100 times thicker than the dayside atmosphere of Io (1--4\,nbar, e.g. \citealt{dePater2023}). Given that the inferred surface pressure of \target\ lies well above the $\sim$1\,pbar threshold below which the atmosphere becomes effectively collisionless, this discovery marks only the second detection (after Makemake; \citealt{Protopapa2025}) of what may be considered a true atmosphere (as opposed to an exosphere) in the outer solar system beyond Pluto. Yet, from thermophysical considerations, only the most massive TNOs have sufficient surface gravity to support stable atmospheres. For objects smaller than 1000--1500\,km in diameter, atmospheric escape is expected to deplete hypervolatile species on timescales much shorter than the age of the solar system \citep[e.g.,][]{Schaller2007,Stern2008,Young2020, Protopapa2025}. \citet{Arimatsu2026} calculated a characteristic escape timescale of $10^{2-3}$\,yr for \target's atmosphere, implying rapid volatile loss. 

In this context, the atmospheric detection on \target\ may indicate a transient phenomenon, possibly associated with a recent impact event or ongoing cryovolcanism \citep{Arimatsu2026}. The likely transient nature of its atmosphere, combined with the object’s substantially lower surface gravity compared to the dwarf planets (0.1\,\mss\ vs. 0.4\,\mss\ for Makemake and 0.61\,\mss\ for Pluto), places \target\ in a hitherto unexplored regime for TNO atmospheric studies, providing a valuable testing ground for our broader understanding of cold, tenuous atmospheres.

In this Letter, we present an analysis of JWST spectroscopic observations of \target\ obtained in 2022 and 2024. Motivated by the recent occultation result, we derive quantitative constraints on the presence and abundances of the key volatile species \methane\ and CO through comparisons with synthetic fluorescence models. Our results will serve as an important reference point for future efforts aimed at understanding the physical nature, composition, and temporal evolution of \target's atmosphere.

\section{Observations and Data Reduction}\label{sec:obs}

\begin{figure*}[t]%
\centering
\includegraphics[width=0.9\textwidth]{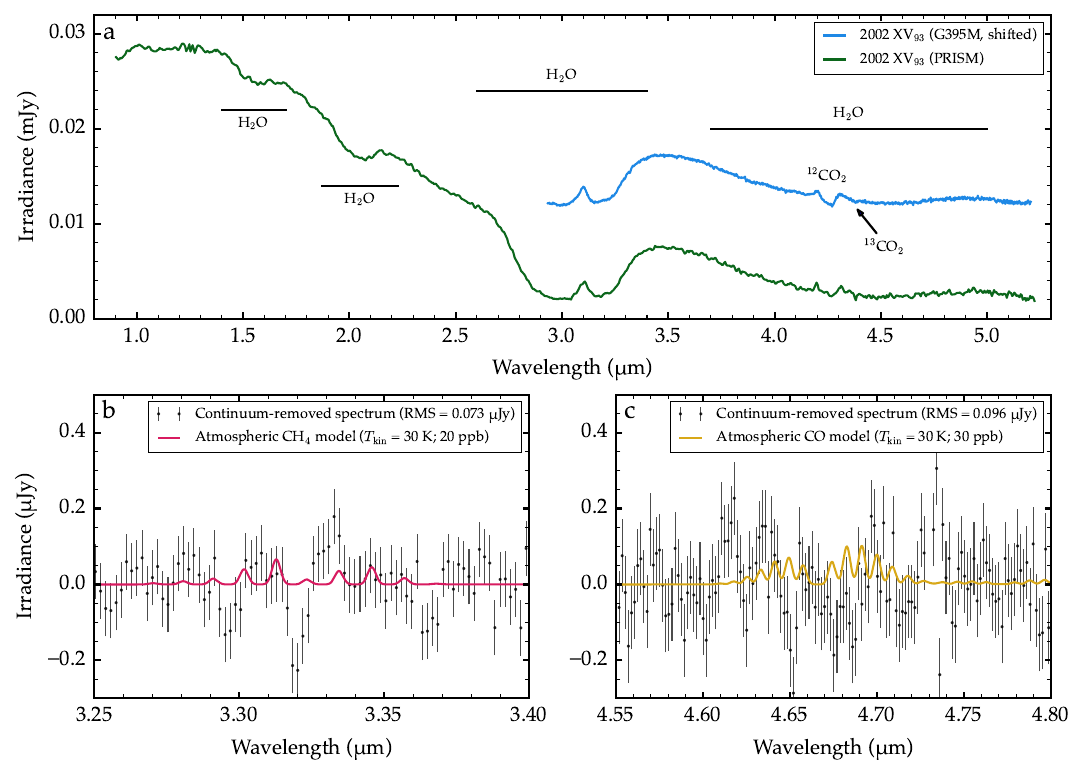}
\caption{Panel~(a) JWST/NIRSpec IFU irradiance spectra of \target\ obtained from observations with the PRISM and G395M dispersers and extracted using a $0\overset{''}{.}6$-diameter circular aperture. The G395M spectrum is offset by 0.01\,mJy for clarity. Major absorption features of \water\ and \cotwo\ are labeled. Panel~(b) Continuum-removed spectrum in the \methane\ fluorescence region (black points). The per-point uncertainties are set to the rms value of the residuals, which is reported in the legend. The magenta curve shows the synthetic atmospheric \methane\ spectrum that was generated assuming a surface pressure of 160\,nbar and an isothermal gas kinetic temperature of 30\,K. The mixing ratio listed in the legend is the $1\sigma$ upper limit and was determined by matching the maximum modeled emission amplitude to the calculated residual rms value. Panel~(c) Same as panel~(b), but for the CO fluorescence region.}\label{fig:1}
\end{figure*}

\target\ was observed with the integral field unit (IFU) on JWST's Near Infrared Spectrograph (NIRSpec) during two separate epochs. The first observation was obtained using the double-pass PRISM disperser on UT 2022 November 3 as part of Cycle 1 Program \#2418 (PI: N.~Pinilla-Alonso; \citealt{PinillaAlonso2025}), producing a continuous spectrum from 0.6 to 5.3\,\um\ at a spectral resolution of $R\,\sim\,30$--300. The second observation occurred on UT 2024 October 8 as part of Cycle 3 Program \#4627 (PI: M.~Brown) and utilized the medium-resolution G395M grating with the F290LP filter, which provides coverage of the 2.9--5.1\,\um\ region at $R\,\sim\,1000$. Both observations employed a four-point dither pattern to mitigate the effects of cosmic rays and improve the spatial sampling of the target's point-spread function (PSF). \autoref{tab:obs} lists the observational details of the two JWST visits. For comparison, the viewing geometry of the UT 2024 January 10 occultation measurements is also included.

The publicly accessible data from the JWST observations were downloaded from the Mikulski Archive for Space Telescopes (MAST) and analyzed using the \texttt{jwstspec} tool \citep{jwstspec}.\footnote{Detailed descriptions of the capabilities of \texttt{jwstspec} can be found in several published works \citep[e.g.,][]{emery2024,wong2024}.} The stacks of uncalibrated up-the-ramp detector readouts were passed through the first two stages of the official JWST calibration pipeline (Version 2.0.1; \citealt{jwst}) to produce dark-corrected, flat-fielded, flux-calibrated, and spatially rectified IFU data cubes for each dithered exposure. The $1/f$ detector readnoise systematic was removed from the intermediate count-rate images using the optional \texttt{clean\_flicker\_noise} step. To examine the target's PSF profile in detail, we also generated dither-combined data cubes using Stage 3 of the calibration pipeline, with the \texttt{outlier\_detection} step skipped to prevent spurious flagging of pixels near the target centroid during the resampling process.

We carried out custom spectral extraction on the individual dithered exposures by applying the specialized empirical PSF fitting method contained in \texttt{jwstspec}. Each wavelength slice was fit with a PSF template derived from the data cube using a moving-median filter with a width of 21 wavelength elements. The spectrum was then measured from the best-fit PSF templates within a 6-pixel-diameter ($0\overset{''}{.}6$) circular aperture centered on the target's calculated centroid position. The background flux level was set as the median pixel value outside of a $21\times21$ pixel box surrounding the target. The individual dither spectra were cleaned of outliers using a 21-point-wide $5\sigma$ moving-median filter. To ensure that the outlier flagging routine was not erroneously masking data points containing possible fluorescence emission from atmospheric molecules, the spectra from the dithered exposures were carefully inspected through comparisons with one another as well as with the spectrum available on MAST extracted from the dither-combined Stage 3 data cubes without outlier rejection. The shapes of the spectra were consistent at all wavelengths across the four dithered exposures. After this vetting procedure, the dither spectra were averaged together to produce the combined irradiance spectrum.

To correct for wavelength-dependent aperture flux losses, calibration curves were derived from JWST observations of the solar-type standard stars SNAP-2 (PRISM; Program \#1128) and GSPC~P330-E (G395M; Program \#6606) that utilized the same dither and readout patterns. Those data were processed in an identical manner to the \target\ observations, and the spectra were extracted using the same aperture size and background region. We then computed the ratio between the measured stellar spectra and the corresponding CALSPEC models \citep{calspec}, smoothed the ratio arrays using a third-order Savitzky--Golay filter, and multiplied the resultant vectors to the \target\ spectra. To remove regions of low throughput and poor data quality, the PRISM spectrum was trimmed to exclude wavelengths below 0.9\,\um. The final irradiance spectra of \target\ from the two JWST observations are shown in panel~(a) of \autoref{fig:1}.


\section{Modeling and Discussion}\label{sec:disc}

The low-resolution PRISM spectrum of \target\ displays major absorption bands of \water\ ice at 1.5, 2.0, 3.0, and 4.5\,\um, as well as a prominent Fresnel reflectance peak at 3.1\,\um. The absorption band at $\sim$4.25\,\um\ is attributed to the fundamental $\nu_3$ asymmetric stretching mode of solid \cotwo. The medium-resolution G395M spectrum reveals the morphology of this \cotwo\ feature in greater detail, showing enhanced reflectance shoulders above the local continuum that bracket a strongly asymmetric absorption centered near 4.27\,\um. At 4.38\,\um, the $\nu_3$ band of the $^{13}$\cotwo\ isotopologue is also discernible. Overall, the spectral profile of \target\ is consistent with the broader \water-rich class of TNOs (often alternatively referred to as ``bowl''-type TNOs) that exhibit prominent \water-ice features---one of the three major compositional types that have been identified from ensemble analyses of JWST TNO spectra \citep{Holler2025RNAAS,PinillaAlonso2025,Wong2025}.

At the low temperatures of TNO surfaces ($\lesssim$50\,K), only the most volatile molecules---\nitrogen, \methane, and CO---and their photochemical products can persist in the gas phase and sustain a thin atmosphere. We probed the G395M spectrum of \target\ for the spectral signatures of gaseous volatiles, focusing on \methane\ and CO. Under the non-local thermodynamic equilibrium (non-LTE) conditions expected in rarefied atmospheres, \methane\ and CO are predicted to fluoresce following excitation by solar radiation, producing characteristic emission features superimposed on the continuum that would be resolvable at the spectral resolution of the G395M grating. Although \nitrogen\ may be the dominant species in the atmosphere, its near-infrared features are limited to collision-induced absorption in the fundamental 4.3\,\um\ band and its overtone at 2.15\,\um \citep[e.g.,][]{Lafferty1996}, resulting in negligible opacities in submillibar atmospheres. Moreover, the \nitrogen\ fundamental band coincides with the \cotwo-ice absorption feature on the surface (see \autoref{fig:1}, panel~(a)). It would therefore be challenging to disentangle any possible weak \nitrogen\ emission lines from the complex underlying solid-state spectral profile.

The strongest \methane\ fluorescence band in the near-infrared is centered at 3.31\,\um\ and corresponds to the $\nu_3$ asymmetric stretching mode. To search for the spectral signature of atmospheric \methane, we modeled the 3.25--3.40\,\um\ continuum with a cubic polynomial function and subtracted the best-fit trend from the irradiance spectrum. The continuum-removed spectrum in the \methane\ fluorescence region is shown in panel~(b) of \autoref{fig:1}. The computed rms metric of the residual array---0.073\,\uJy---quantifies the intrinsic scatter of the data across this wavelength range. The residuals exhibit significant correlated noise, and we therefore adopted the rms value as an empirical estimate of the $1\sigma$ detection limit for atmospheric emission features. In the case of CO, the fundamental $v=1$ stretching band is located near 4.7\,\um. We removed the local continuum using a cubic polynomial fit across the 4.55--4.80\,\um\ region. The resulting residual spectrum is shown in panel~(c) of \autoref{fig:1} and has an rms value of 0.096\,\uJy.

To derive constraints on the abundance of \methane\ and CO in \target's atmosphere, we compared the measured continuum-subtracted irradiance to synthetic atmospheric model spectra, following a similar semiempirical approach to the one used to interpret the \methane\ $\nu_3$ fluorescence band detected in Makemake's JWST/NIRSpec G395M spectrum \citep{Protopapa2025} and the \methane\ $\nu_4$ fluorescence in Pluto's JWST MIRI spectrum \citep{Lellouch2025}. Unlike \citet{Protopapa2025}, who also explored an expanding coma scenario for Makemake, we restricted our analysis to a gravitationally bound atmosphere, following the interpretation of \citet{Arimatsu2026}. We considered isothermal atmospheres at different kinetic temperatures ($\tkin$) and incorporated prescribed vibrational temperature ($\tvib$) profiles into an otherwise LTE radiative transfer framework initially developed by \citet{Lellouch2022} using the formalism of \citet{Edwards1993}. We adopted a representative surface pressure of 160\,nbar, consistent with the range of best-fit pressures inferred from the atmospheric refraction analysis of the 2024 stellar occultation event \citep{Arimatsu2026}, and adjusted the \methane\ and CO mixing ratios, which were assumed to be vertically uniform.


While Makemake's atmosphere may be in hydrostatic equilibrium, with pressure decreasing exponentially with altitude $z$, this is almost certainly not the case for \target, given the object's small size ($R\,\sim\,255$\,km) and correspondingly weak surface gravity. As pointed out by \citet{Arimatsu2026}, the Jeans parameter of \target's atmosphere is close to unity. It follows that the hydrodynamic state of the atmosphere is expected to approach the free molecular flow regime, with the density decreasing as $r^{-2}$, where $r\,=\,R\,+\,z$ \citep[e.g.,][]{Volkov2011}. We therefore adopted the $r^{-2}$ density profile and considered three different gas $\tkin$ values---30, 70, and 100\,K. The first temperature represents an end-member cold atmosphere that is unaffected by radiative processes, while the other two are typical of Pluto's atmospheric temperatures in the upper atmosphere and near the stratopause. The $r^{-2}$ density profile entails a large spatial extent for \target's atmosphere, and we tailored the model outputs to the aperture used for extracting the JWST spectrum. At $\Delta\,=\,37.86$\,au, a $0\overset{''}{.}6$-diameter aperture corresponds to $\sim$16,500\,km. When producing disk-integrated synthetic spectra, we calculated the limb-tangent radiances for altitudes up to 50,000\,km and spatially convolved the results using a two-dimensional Gaussian with an FWHM equal to the aperture size.
 
\begin{figure}[t]%
\centering
\includegraphics[width=\columnwidth]{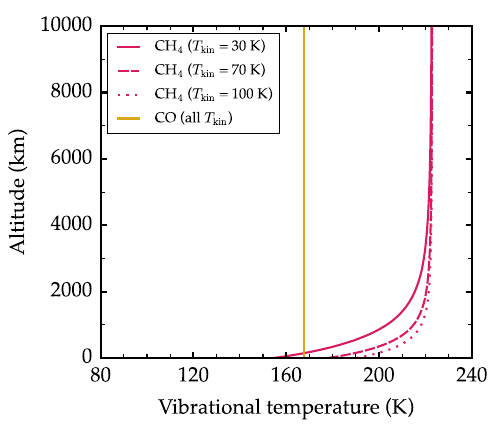}
\caption{Vibrational temperature ($\tvib$) profiles for \methane\ (magenta) and CO (yellow) used in modeling \target's atmosphere. For \methane, the prescribed $\tvib$ profiles at the different assumed gas kinetic temperatures ($\tkin$) show a strong dependency on altitude due to the transition from partial thermal equilibrium at low altitudes to fluorescence equilibrium in the upper atmosphere. In contrast, the isothermal CO $\tvib$ profile is fixed at 167.5\,K for all $\tkin$.}\label{fig:2}
\vspace{-0.3cm}
\end{figure}

To estimate the $\tvib$ profiles, we followed the guidance from the fully self-consistent non-LTE models developed for Triton (I.~Wong~et~al.~2026, in preparation), taking into account the difference in heliocentric distance between Triton (29.9\,au) and \target\ (37.9\,au). We only included the primary \methane\ $\nu_3$ and CO(1--0) bands, as the combination and hot bands (e.g., \methane\ $\nu_3\,+\,\nu_4\,-\,\nu_4$ and CO(2--1)) are expected to be much weaker. Under fluorescence equilibrium---a reasonable assumption in the upper part of the atmosphere---the expected $\tvib$ for the \methane\ $\nu_3$ and CO(1--0) levels are 5.5 and 4.5\,K colder at \target\ compared to Triton, respectively. Thus, for the CO(1--0) band, we based the $\tvib$ profile on the roughly isothermal value of 172\,K calculated for Triton (I.~Wong~et~al.~2026, in preparation) and adopted an altitude-independent $\tvib$ of 167.5\,K. For \methane, the non-LTE models of Triton's atmosphere indicate that $\tvib$ transitions from fluorescence equilibrium in the upper atmosphere to a monotonically decreasing trend in the denser lower atmosphere, where $\tvib$ becomes influenced by $\tkin$ and the departure from LTE is pressure dependent. The prescribed  \methane\ and CO $\tvib$ profiles are shown in \autoref{fig:2}. 

Using the semiempirical approach adopted by \citet{Lellouch2025} and \citet{Protopapa2025}, we generated synthetic spectra for the three $\tkin$ cases discussed above. Neither \methane\ nor CO emission bands are discernible in the continuum-subtracted spectra. To derive $1\sigma$ upper limits on the molecular abundances, we adjusted the \methane\ and CO mixing ratios such that the maximum emission amplitude matched the residual rms values for the continuum-subtracted spectra. To illustrate the predicted fluorescence band profiles, the $\tkin\,=\,30$\,K model spectra are shown in panels~(b) and (c) of \autoref{fig:1}. Across the three $\tkin$ cases, we obtained a range of stringent $1\sigma$ upper limits on the \methane\ and CO mixing ratios: 20--60 and 30--200\,parts per billion, respectively, which correspond to vertically integrated column densities of $(1.6-2.0)\,\times\,10^{13}$\,cm$^{-2}$ and $(3-6)\,\times\,10^{13}$\,cm$^{-2}$ and surface partial pressures of $(3-10)\,\times\,10^{-6}$ and $(5-30)\,\times\,10^{-6}$\,nbar.

While one could consider developing physical non-LTE excitation models for \target's atmosphere, including radiative and collisional processes, such modeling is unlikely to significantly refine the upper limit mixing ratio constraints derived from the presented nondetections. Although there is some inherent uncertainty in the adopted $\tvib$ profiles shown in \autoref{fig:2}, its impact on the reported upper limits is minor. As a test, we considered perturbations to the CO $\tvib$ profile of $\pm$5\,K at all altitudes. These modified $\tvib$ profiles changed the derived abundance upper limits by at most a factor of 2. A similar result was obtained for \methane\ when 20 and 5\,K uncertainties in the $\tvib$ profile were applied at the surface and at an altitude of 10,000\,km, respectively. We therefore conclude that plausible uncertainties in the adopted $\tvib$ profiles have only a limited effect on the inferred abundance constraints.

Our modeling has constrained the abundance of \methane\ and CO in \target's atmosphere to at most trace amounts. It follows that the bulk gas composition must be dominated by other species that do not show strong fluorescence signatures in the observed wavelength range, such as \nitrogen\ or a noble gas (e.g., argon). However, invoking such hypervolatile compounds exacerbates the issue of the extremely short atmospheric depletion timescale from Jeans escape. An alternative explanation is that \target\ could actually host a \methane-dominated atmosphere, but the gas is restricted to a near-surface layer, with the density falling off steeply with altitude due to insolation-driven \methane\ photolysis. This scenario has been invoked to explain the lack of fluorescent \methane\ emission on Triton, even though \methane\ is known to be present in Triton's atmosphere and is detected in absorption at near-infrared wavelengths (\citealt{Lellouch2010}; I.~Wong~et~al.~2026, in preparation). In this case, some replenishment mechanism would be necessary to maintain a sufficiently high concentration of \methane\ gas to account for the surface pressure inferred from the stellar occultation measurement.

\begin{figure}[t]%
\centering
\includegraphics[width=\columnwidth]{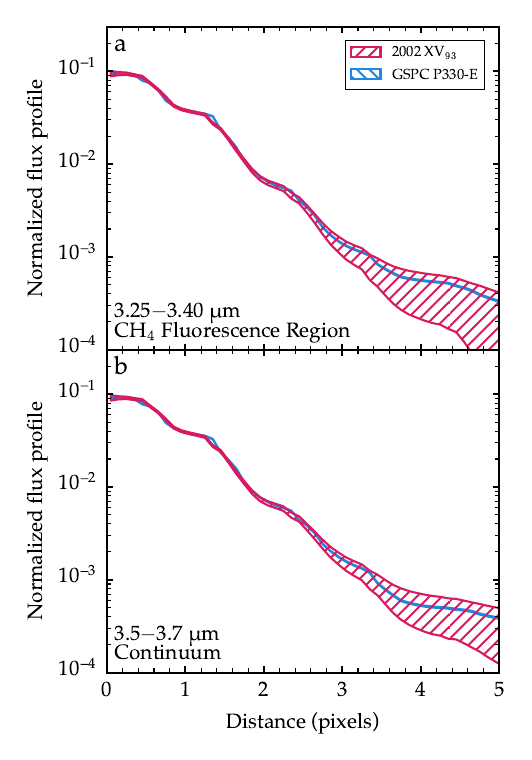}
\vspace{-0.2cm}
\caption{Panel~(a) Comparison of the normalized radial flux profiles of \target\ (magenta) and the solar-type standard star GSPC~P330-E (blue), averaged across the 3.25--3.40\,\um\ wavelength range that contains the \methane\ $\nu_3$ fluorescence band. Panel~(b) Same as panel~(a), but for the 3.5--3.7\,\um\ continuum region. In both cases, the PSF of \target\ is consistent with that of a point source at all separations, indicating the nondetection of extended emission.}\label{fig:3}
\vspace{-0.2cm}
\end{figure}

The hydrodynamical regime of gaseous volatiles in \target's atmosphere raises the possibility of significant outflow, which may entrain solid grains. This scenario is analogous to cometary comae, where sublimation from the surface or subsurface expels dust and/or ice, leading to spatially extended emission around the central nucleus. In the case of \target, diffuse solid-state material would contribute to the observed occultation signature and could therefore lower the density of the gaseous atmosphere necessary to reproduce the ingress and egress light curves. To search for evidence of extended emission, we examined the shape of \target's PSF in the calibrated dither-combined images. We computed the azimuthally averaged radial profiles in two wavelength ranges: (1) 3.25--3.40\,\um, which spans the region of \methane\ fluorescence, and (2) 3.5--3.7\,\um, which samples the nearby continuum and is sensitive to ejected grains. These profiles were compared to analogously derived radial profiles of the solar-type standard star GSPC~P330-E. As illustrated in \autoref{fig:3}, the spatial morphology of \target's PSF is indistinguishable from that of a point source at all separations, indicating the nondetection of both a \methane\ coma and extended emission from solid-state material. It is important to note, however, that due to the large heliocentric distance of \target\ and the corresponding coarse spatial resolution of the NIRSpec IFU data cubes ($\sim$2,750\,km\,pixel$^{-1}$), we cannot entirely exclude the possibility of extended structures that are confined close to the surface.

\vspace{-0.4cm}
\begin{acknowledgments}
This work is based on observations made with the NASA/ESA/CSA James Webb Space Telescope. The data were obtained from the Mikulski Archive for Space Telescopes at the Space Telescope Science Institute, which is operated by the Association of Universities for Research in Astronomy, Inc., under NASA contract NAS 5-03127 for JWST. These observations are associated with Programs \#2418 and \#4627. The specific observations analyzed can be accessed via \dataset[https://doi.org/10.17909/kd4j-da74]{https://doi.org/10.17909/kd4j-da74}
\end{acknowledgments}

\vspace{-0.15cm}
\facilities{JWST.}
\vspace{-0.15cm}
\software{\texttt{astropy} \citep{astropy2013,astropy2018,astropy2022}, \texttt{jwst} \citep{jwst}, \texttt{jwstspec} \citep{jwstspec}, \texttt{matplotlib} \citep{matplotlib}, \texttt{numpy} \citep{numpy}, \texttt{scipy} \citep{scipy}.}


\vspace{-0.15cm}
{\small\bibliography{main}{}}
\bibliographystyle{aasjournal}

\end{document}